# Size Effects in Nanocrystalline Thoria

*Tatiana V. Plakhova,[1] Anna Yu. Romanchuk,[*,1] Daria V. Likhosherstova [1] Alexander E. Baranchikov,[1,2] Pavel V. Dorovatovskii,[3] Roman D. Svetogorov,[3] Tatiana B. Shatalova,[1] Tolganay B. Egorova,[1] Alexander L. Trigub,[3] Kristina O. Kvashnina,[4,5] Vladimir K. Ivanov,[2] and Stepan N. Kalmykov[1].*

1 - Lomonosov Moscow State University, Department of Chemistry, Moscow, Russia

2 - Kurnakov Institute of General and Inorganic Chemistry, Russian Academy of Sciences, Moscow, Russia

3 - National Research Centre "Kurchatov Institute", Moscow, Russia

4 Rossendorf Beamline at ESRF – The European Synchrotron, Grenoble, France

5 - Helmholtz Zentrum Dresden-Rossendorf (HZDR), Institute of Resource Ecology, Dresden, Germany

*corresponding author: romanchuk.anna@gmail.com, phone number: +7-495-939-3220

ABSTRACT

The facile chemical precipitation method and subsequent thermal treatment were shown to be suitable for preparation of crystalline $ThO_2$ nanoparticles (NPs) in a wide range of particle sizes (from 2.5 to 34.3 nm). The obtained NPs were investigated with X-ray diffraction, high-resolution transmission electron microscopy and X-ray absorption techniques to find out the possible size effects associated with nanocrystalline thoria. For 2.5 nm NPs, the lattice parameter of $ThO_2$ was found to increase by up to 1.1 %, in comparison with the bulk material. The decrease in the particle size was also accompanied by a significant decrease in the Th-Th coordination number.

INTRODUCTION

The nanoscale approach is state-of-the-art in the field of advanced materials design [1-4]. The basis of this approach is the change in material properties with a decrease in the size of structural elements (grains, phase inclusions, layers, pores, etc.) to 1 - 100 nm. With a decrease in the size of a single "building block" [3], the substance's mechanical, optical, catalytic and conductive properties could be changed.

In recent years, actinide-based nanoparticles have attracted a great deal of attention in the field of environmental safety control and in the development of new technological schemes [5-8]. Among other actinide oxides, thorium dioxide possesses excellent chemical stability and a high melting point. This makes thorium dioxide a material of choice for catalyst preparation [9-11]. Alternative uses of thoria, such as ceramics, solid electrolytes and optical materials, have also been reported [12-15]. Currently, the most promising area of $ThO_2$ application (including in a nanosized state) is considered to be the nuclear industry [16]. The amount of thorium in the Earth's crust is three to four times greater than that of uranium, so the development of a nuclear fuel cycle based on thorium-232 as the fertile material is inherent in the nuclear policy of a number of countries, including India, Russia, China etc. The development of a synthetic pathway for thoria nanoparticles (NPs) preparation with controlled size and properties is of crucial importance for developing this concept, since an important processing goal for nuclear ceramics is to obtain nuclear fuel pellets with a uniform microstructure and with a pre-determined grain size. [17] Moreover, according to recent studies, the grain boundaries and interfaces may serve as effective sinks for defects caused by radiation damage. Thus, nanocrystalline materials containing a larger fraction of such defects could improve the radiation resistance of fuel pellets in comparison with their microcrystalline counterparts [18-19].

To date, the following methods have been proposed for thoria NPs preparation: thermal decomposition of thorium salts; [20-25] the hydrothermal route; [21, 26-27] template synthesis; [28] electrochemical deposition; [29-30] the sol-gel technique [31-32] and the "heating-up" method. [33] The thermal decomposition method of $ThO_2$

NPs production is the most widely used. In a series of papers by Dash et al., thorium nitrate pentahydrate, [20] thorium oxalate hexahydrate [22] and thorium carbonate [25] were used as the precursors for $ThO_2$ synthesis. It was found that the crystallization of $ThO_2$ from amorphous and anhydrous precursors starts at about 380 °C. The primary nanocrystals formed at this temperature have a size of about 2.5 - 3.0 nm. Further size tuning can be achieved by setting up appropriate annealing conditions: an increase in the annealing time and temperature leads to an increase in crystallite sizes. The use of organic chelating agents makes it possible to obtain crystalline $ThO_2$ NPs with well-controlled morphology.[27, 31-33] For the first time, Hudry et al. [33] synthesized rod-like particles of thorium dioxide using the so-called "heating-up" method. Yousefi et al. [29-30] proposed an electrochemical method for obtaining nanocrystalline thorium dioxide with a high surface area by annealing primary precipitate formed from thorium nitrate by electrochemical deposition.

The chemical precipitation technique is a versatile method for crystalline nanomaterials synthesis, even at room temperature, allowing for facile control of the size and morphology of the products by varying the composition of initial solutions and their mixing parameters. Moreover, the chemical precipitation technique is of great interest for environmentally benign science mass production, as it does not require any special conditions or costly equipment. Chemical precipitation of crystalline $ThO_2$ particles has been reported by several researchers. [28, 34] In these papers, thermal treatment was used to enable the formation of highly crystalline $ThO_2$ particles. Batuk et al. [28] precipitated $ThO_2$ in a mesoporous matrix at room temperature and with further thermal treatment at 250 ºC in air. By varying the pH values of the initial suspension, crystalline $ThO_2$ NPs of 2, 3 and 5 nm size were produced. Dzimitrowicz et al. [34] succeeded in obtaining crystalline $ThO_2$ NPs of 3 nm size by precipitation from an aqueous thorium nitrate solution, with an excess of KOH at 0 °C and then further vacuum drying at 20 °C. Thus, only fragmentary information is still available in the literature concerning the formation of crystalline $ThO_2$ NPs during chemical deposition and subsequent thermal treatment, and a simple size-tuneable synthesis of $ThO_2$ NPs with narrow size distributions from aqueous systems is still a great challenge.

The influence of particle size on thorium oxide properties is poorly discussed in the literature. Existing examples include the effects of $ThO_2$ NPs' size on its solubility [35], adsorption properties [27] and toxicity [36]. The solubility and adsorption capacity of $ThO_2$ NPs increase noticeably with a decrease in particle size. The origins of the phenomena are typically associated with the high surface area of nanoparticles, being strongly dependent on particle size, especially for 1 - 10 nm NPs. He et al. [36] investigated the aquatic toxicity of $ThO_2$ NPs and indicated that NPs with a size of 50 nm were more toxic to the algae than 140 nm NPs. They also explained this fact by the increase in surface area, since smaller $ThO_2$ NPs should have more active surface sites, compared to larger ones, and thus produce more toxic reactive oxygen species in the cells. Nevertheless, the exact nature of the size effects on thoria nanoparticles remains unexplored.

The present study is dedicated to the controlled synthesis, the structural characterization and the investigation of the size-related structural features of thorium oxide NPs. To achieve this goal, an approach based on the sequential heat treatment of freshly precipitated particles was proposed. Gentle heating in air, and hydrothermal treatment or high temperature annealing of dried samples, were used for $ThO_2$ NPs preparation in a wide range of sizes (2.5 - 34 nm). By carefully adjusting heat treatment conditions such as time, temperature and the composition of the medium (for hydrothermal treatment), the size of $ThO_2$ NPs can be adequately controlled. The resulting nanoparticulate $ThO_2$ samples were characterized using advanced analytical techniques, including high-resolution transmission electron microscopy (HRTEM) coupled with electron diffraction (ED), and X-ray spectroscopic methods. In this paper, the influence of $ThO_2$ crystallite size on key structural characteristics such as the local atomic structure and the unit cell parameters was determined. The full-range dependence of the $ThO_2$ lattice parameter on particle size was revealed for the first time.

# EXPERIMENTAL SECTION

## Synthesis procedure

Thorium oxide samples were synthesized by the following methods. Aqueous solutions of thorium nitrate pentahydrate (0.1 M, 0.5 M or 1 M) were rapidly added, under vigorous stirring, to aqueous ammonia (3 M) or sodium hydroxide (3 M) aqueous solutions, both taken in excess. The mixtures were stirred continuously for an hour under ambient conditions. The precipitates were collected and thoroughly washed with MilliQ water. Before characterization, all the samples were dried at 40 or 150 °C in air.

For hydrothermal treatment, the as-prepared sample precipitated from 0.1M $Th(NO_3)_4 \cdot 5H_2O$ aqueous solution using 3M NaOH aqueous solution was used as the precursor. Hydrothermal processing was carried out in MilliQ water, or in 3M NaOH solution, at 120 °C, 150 °C, 180 °C or 210 °C for 24 hours. After hydrothermal treatment, the samples were also dried for further characterization. For annealing experiments, a thoria sample synthesized from 1M $Th(NO_3)_4 \cdot 5H_2O$ and 3M $NH_3 \cdot H_2O$ and dried at 150 °C was used. High temperature annealing experiments were conducted in a muffle furnace at 300 °C, 400 °C, 500 °C, 600 °C and 700 °C for 1 to 8 hours in air.

## Characterization of NPs samples

X-ray diffraction (XRD) patterns used for structure analysis were obtained using laboratory and synchrotron X-ray sources. For laboratory measurements, a Bruker D8 Advance diffractometer with a Cu Kα radiation source (wavelength 1.54 Å) was used. The analysis was carried out with a step of 0.02 °2θ, with an accumulation time of 0.3 seconds per point in the range of 10-120 °2θ. For synchrotron-based XRD measurements, a MarCCD165 detector was used. Diffraction patterns were obtained using monochromatic radiation with a wavelength of $\lambda$ = 0.802575 Å (photon energy E = 15448.3 eV), focused on a sample of up to 400 microns in size. Two-dimensional diffraction patterns obtained were further transformed to dependence of the intensity on the scattering angle using Fit2D and Dionis software.

The Scherrer equation was used to determine the crystallite size of the $ThO_2$ NPs from the full width at half maximum (FWHM) of the (111) and (200) diffraction lines**.** The coefficient of anisotropy was set to 1. Line profiles for selected reflexes were fitted to pseudo-Voigt functions. Crystallite size values calculated from full width at half maximum (FWHM) of (111) and (200) diffraction lines differed by no more than 0.2 nm for all the samples. Instrumental broadening was taken into account when calculating the particle size by direct subtraction from the FWHM values. Instrumental broadening for the laboratory XRD equipment was considered to be $0.09 \pm 0.01$ °2θ. For the synchrotron based XRD, instrumental broadening was calculated using the Caglioti equation, [37] based on standard calibrations. The calibration of the angular scale of the detector and determination of the hardware broadening of diffraction reflexes were carried out by measuring the diffraction pattern of the polycrystalline standard $Na_2Ca_3Al_2F_{14}$ (NAC NIST SRM).

For a full-profile analysis of the diffraction patterns, and further unit cell parameter calculation, TOPAS 4.2 software was applied. Fourth-order Chebyshev polynomials with a reciprocal term were used to fit the background. The overall fitting was performed using the fundamental parameter approach.

Microstructure evaluation was performed using a Carl Zeiss Libra 200 and Jeol-2100F high-resolution transmission electron microscopes (HRTEMs) at an accelerating voltage of 200 kV. Electron diffraction patterns were recorded using the same instruments. To estimate average particle size and distribution from HRTEM images, sets of > 200 particles were used.

The study of the thermal behaviour of the samples and the composition of the gaseous products evolved upon heating was carried out on a NETZSCH STA 409 PC Luxx synchronous thermal analyzer, in combination with a NETZSCH QMS 403 C Aëolos quadrupole mass spectrometer. Samples were heated in a dynamic air flow (30 mL/min) to a temperature of 1000 °C at a rate of 10 °C/min. The analysis of thermograms and mass spectra was performed using NETZSCH Proteus software.

In the present investigation, we used XAS spectroscopy to analyze Th local atomic geometry in $ThO_2$ NPs. The Th $L_{III}$-edge XANES and EXAFS spectra were recorded at the Rossendorf Beamline BM20 [38] of the European Synchrotron (ESRF) (Grenoble, France) and the Structural Materials Science beamline [39] of Kurchatov Synchrotron Radiation Source (Moscow, Russia). At ESRF, the incident energy was selected using the <111> reflection from a double water-cooled Si crystal monochromator. Rejection of higher harmonics was achieved using two Rh mirrors at an angle of 2.5 mrad relative to the incident beam. The incident X-ray beam had a flux of approximately $2\cdot10^{11}$ photons $s^{-1}$ on the sample position. XAFS data were recorded in fluorescence mode using a 13-element high-throughput Ge-detector. The recorded intensity was normalized to the incident photon flux. Data were collected up to $k = 10$ $A^{-1}$, with a typical acquisition time of 20 min per spectrum. At the Kurchatov Synchrotron Radiation Source, the storage ring with an electron beam energy of 2.5 GeV and a current of 80–100 mA was used as the source of radiation. For the monochromatization of the X-ray beam the Si(111) channel-cut monochromator was used, which provided an energy resolution of $\Delta E/E \sim 2\cdot10^{-4}$. Dumping of higher energy harmonics was achieved by distortion of the monochromator geometry. Energy calibration was performed by measuring XAS spectra of Fe foil. All experimental data were collected in transition mode, and intensities of incident and transmitted X-ray beams were measured using ionization chambers filled with $N_2$. In the measurements of XAS spectra, three ionization chambers were used, providing simultaneous measurements of XAS spectra for the sample and reference. In this way, the energy calibration of the monochromator could be checked and corrected. At every energy point in the XANES region, the signal was integrated for 1 second; in the EXAFS region, integration time was set to 1 second at the beginning of the region and increased to 4 seconds at the end of the spectra. For all the samples, at least three experimental spectra were collected and merged using IFEFFIT software.[40]

EXAFS data ($\chi_{exp}(k)$) were analyzed using the IFEFFIT data analysis package. EXAFS data reduction used standard procedures for pre-edge subtraction and spline background removal. The Fourier transformation (FT) of the $k^3$-weighted EXAFS functions $\chi_{exp}(k)$ was calculated over the ranges of photoelectron wave numbers k = 3–11.0 Å$^{-1}$. Fitting of the EXAFS spectra were performed in R-space by varying the structural parameters, including interatomic distances ($R_i$) and coordination numbers ($CN_i$). Debye–Waller factors ($\sigma^2$) were found by the non-linear fitting of theoretical spectra (Equation (1)) to experimental spectra.

$$\chi(k) = S_0^2 \sum_{i=1}^{n} \frac{CN_i F_i(k)}{R_i^2 k} e^{\frac{-2R_i}{\lambda(k)}} e^{-2\sigma_i^2 k^2} sin\left(2kR_i + F_i(k)\right) \qquad (1)$$

The theoretical data were simulated using the photoelectron mean free path $\lambda(k)$, amplitude $F_i(k)$ and phase shift $F_i(k)$ calculated ab initio using program FEFF6.[41] For the EXAFS spectra fitting, the Debye-Waller factor for the Th-Th scattering path was set at 0.005 Å$^{-2}$, this value being optimal for the larger $ThO_2$ particles. The $S_0^2$ parameter value was set at 0.9. The intensities of scattering paths were calculated taking into account the crystal structure of $ThO_2$.[42]

## RESULTS AND DISCUSSION

To determine the conditions for the synthesis of $ThO_2$ NPs of controlled size, we investigated the influence of thorium nitrate stock solution concentration in the range 0.1 - 1M, the type of precipitating agent ($NH_3 \cdot H_2O$ vs. NaOH) and its concentration on the size of NPs. According to XRD data and HRTEM images, crystalline thoria NPs were formed directly under chosen experimental conditions. The changes in the concentration of the initial $Th(NO_3)_4 \cdot 5H_2O$ solution from 0.1M to 1M were found to have no effect on the size of thoria NPs. According to XRD data, thoria precipitation with a 3M ammonia solution, in all cases, provided particles of about 2 nm (Fig. 1). Interestingly, the choice of precipitant could affect the particle size significantly. When using 3M NaOH for $ThO_2$ precipitation, NPs of 3.6 nm size were formed. By changing the NaOH concentration to 1M, 3 nm $ThO_2$ NPs were produced. Comprehensive analysis of XRD data for thoria samples obtained from thorium nitrate solutions of the same concentrations revealed that the use of sodium hydroxide as a precipitant promoted the growth of larger $ThO_2$ NPs.

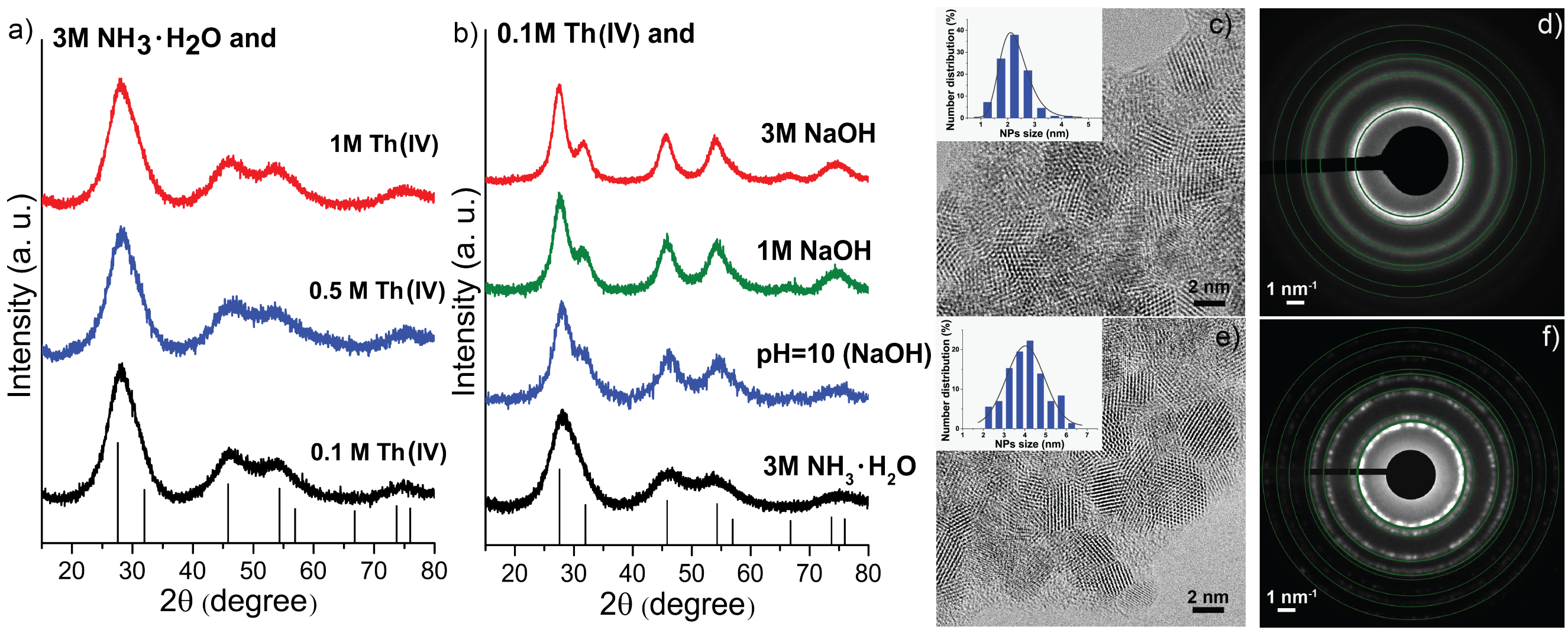

**Figure 1.** a), b) XRD patterns of $ThO_2$ samples precipitated from thorium nitrate solutions of various concentrations (0.1M – 1M) using aqueous ammonia and NaOH aqueous solutions (1M and 3M) as precipitants. HRTEM and ED data for thoria NPs obtained by precipitation from 0.1 M $Th(NO_3)_4$ $5H_2O$ and c), d) 3M $NH_3{\cdot}H_2O$; e), f) 3M NaOH as precipitants. The circles on ED patterns indicate the diffraction pattern calculated for bulk $ThO_2$.[43] The inserts in HRTEM images show the size distributions of $ThO_2$ NPs, as obtained from HRTEM data.

To further elucidate the influence of the precipitant's nature on $ThO_2$ NPs' size, HRTEM measurements were performed for the samples synthesized from 0.1M $Th(NO_3)_4{\cdot}5H_2O$ using 3M $NH_3{\cdot}H_2O$ (Fig. 1c,d) and 3M NaOH (Fig. 1e,f). Electron diffraction patterns confirmed the formation of a cubic $ThO_2$ phase in both samples. The values of the interplanar distances calculated from the electron diffraction data coincide well with previously published data for fluorite type $ThO_2$ (Table 1S). Particle size distributions obtained from HRTEM images are shown in Fig. 1c,e. Both thoria samples had a rather narrow particle size distribution, which can be fitted by a lognormal or Gaussian function. Average NPs diameters were $2.3 \pm 0.2$ nm and $4.0 \pm 0.9$ nm for the samples synthesized with 3M $NH_3{\cdot}H_2O$ and 3M

NaOH, respectively. The crystallite sizes determined from XRD data were in good agreement with those derived from HRTEM.

The effects of hydrothermal treatment temperature and hydrothermal medium composition on $ThO_2$ NPs size were further investigated. The dependence of particle size, as calculated from XRD data, on the temperature and the nature of hydrothermal media is shown in Fig. 2a. The temperature of the hydrothermal treatment had a pronounced effect on particle size: with temperature increase, $ThO_2$ crystallite size increased both in water and NaOH solutions. Again, the particles treated in a sodium hydroxide medium were larger than the particles obtained by treatment in the aqueous medium at the same temperature and with the same duration of treatment. For example, hydrothermal treatment at 210 °C in aqueous media resulted in 4.5 nm thoria NPs, while hydrothermal treatment in 3M NaOH solution under the same conditions produced NPs with an average size of 5.4 nm. According to the literature sources, strongly alkaline conditions promote the formation of anionic forms of $Th^{4+}$ [44], as well as $Zr^{4+}$, $Hf^{4+}$, $Pu^{4+}$ and $Sn^{4+}$ [45-48]. The formation of ternary Th(IV) hydroxide-carbonate complexes under similar experimental conditions was also reported [49]. These anionic species increase significantly the solubility of metal oxides. The rate of $ThO_2$ particle growth under hydrothermal conditions by the mechanism of dissolution/crystallization depends strongly on the solubility of the solid phase. Therefore, larger $ThO_2$ particles in 3M NaOH are believed to be caused by higher solubility of thoria on alkaline media.

According to HRTEM and XRD data, the NPs morphology and phase composition did not change upon hydrothermal treatment, even in a strongly alkaline medium. According to XRD patterns, all the samples obtained by hydrothermal treatment consisted of pure cubic $ThO_2$ (Fig. S1). Both initial $ThO_2$ NPs and the particles treated hydrothermally had the same habitus (Fig. 1e and Fig. 2b).

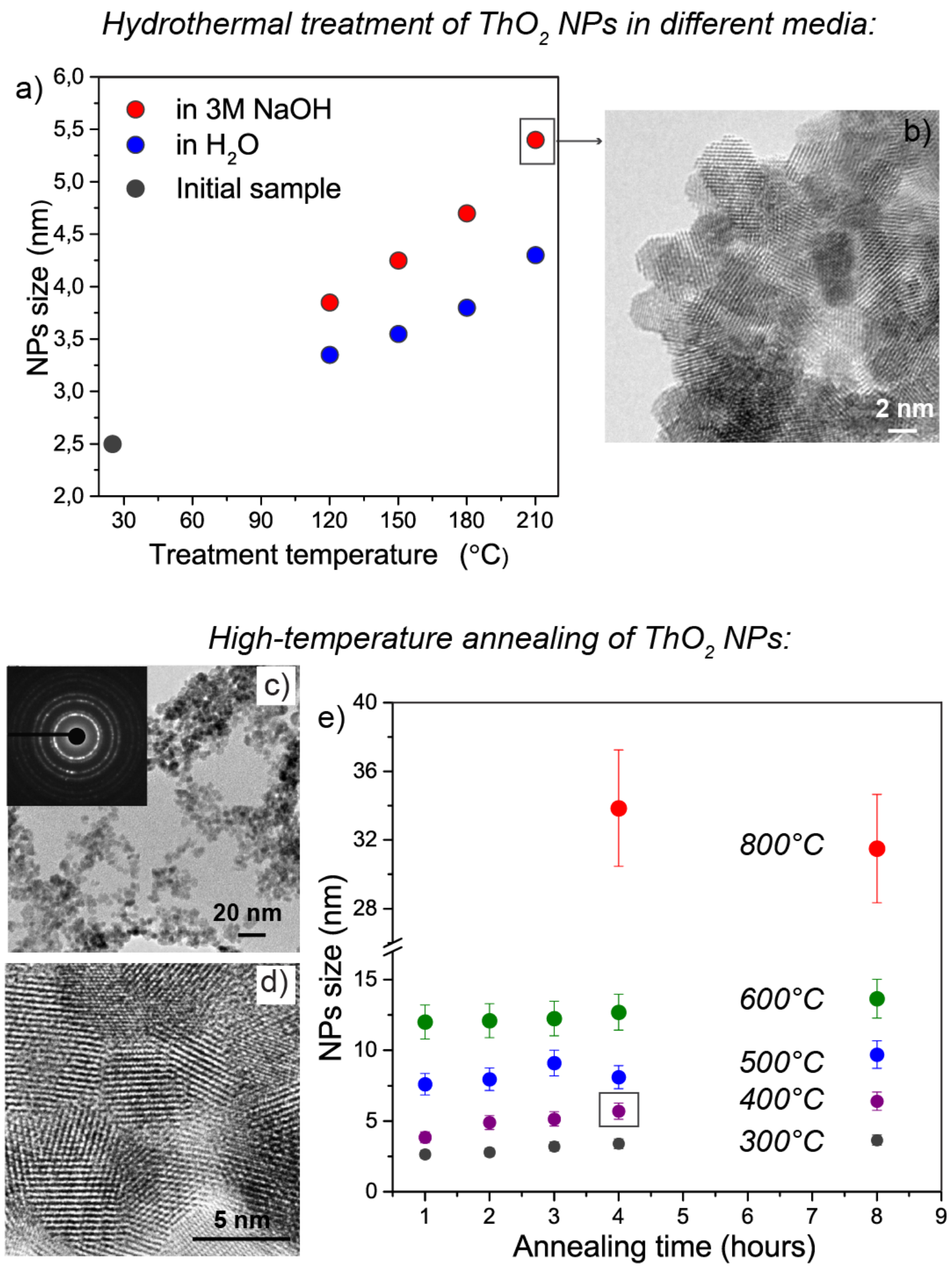

**Figure 2.** a) Dependence of $ThO_2$ NPs size, as calculated from XRD data, on the temperature of hydrothermal treatment in different media. b) HRTEM data for $ThO_2$ sample treated hydrothermally at 210 °C in 3M NaOH. c), d) HRTEM and ED (insert) data for $ThO_2$ sample obtained by annealing thoria NPs at 400 °C for 4 hours. e) Size of $ThO_2$ NPs, as calculated from XRD data, vs. duration and temperature of annealing.

The preservation of NPs' phase composition and morphology during hydrothermal treatment, especially in strongly alkaline media, is not typical for tetravalent metal oxides. For instance, $CeO_2$, which is generally considered a structural analogue of actinide(IV) oxides,[50-53] tends to change its morphology from nearly spherical particles to nanorods, with an increase in the temperature of hydrothermal treatment under strongly alkaline conditions. [54] The same trend is observed for $ZrO_2$, which is also considered to be an analogue of actinide(IV) oxides. Under hydrothermal conditions, the particle shape of $ZrO_2$ changes from being spherical to spindle- or rod-like, with a simultaneous change in phase composition.[55] On the other hand, the formation of crystalline $HfO_2$ in hydrothermal conditions is most similar to $ThO_2$ in terms of phase stability. In particular, according to G. Stefanic et al.,[56-57] hydrous hafnia crystallizes with the formation of m-$HfO_2$ phase, regardless of pH, during the treatment. In the case of $HfO_2$ and $ZrO_2$, the hydrothermal crystallization proceeded much more slowly in neutral pH conditions than in alkaline conditions.[56-57] Similar results were obtained by Montes et al.[58] Note that the same influence of the hydrothermal media was observed for $ThO_2$ NPs growth in the present study.

The formation of larger-size thoria NPs in the presence of sodium hydroxide, both upon direct precipitation from solution and under hydrothermal conditions, could be explained as follows. Generally, the processes of dissolution and recrystallization are the major driving force for NPs growth, both during precipitation from solution at room temperature and under hydrothermal conditions. Larger particles are formed in the presence of the substances that increase their solubility, these substances being often referred to as mineralizers. According to existing data, thorium dioxide solubility has a tendency to increase under high ionic strength conditions.[59] Thus, under the chosen conditions, NaOH solution acts as a mineralizer and contributes to the formation of larger crystalline NPs in comparison to $NH_3 \cdot H_2O$ or aqueous media.

In order to study the relationships between synthesis conditions and particle size in a broader range, experiments on the high temperature growth of $ThO_2$ NPs were carried out. Fig. 2 shows the dependence of the size of NPs, calculated from X-ray diffraction data, on the duration and temperature of annealing. According to the data obtained the crystallite, size increased noticeably with temperature increase. Annealing of the initial sample at 300 – 800 °C for 1 – 8 hours resulted in a systematic increase in NPs size from 2 – 3 to 32 – 34 nm (Fig. S2).

Based on the dependence of the particle size of thorium dioxide on annealing duration, an assumption about the mechanism of crystallite growth during high-temperature annealing can be drawn. A parabolic law describes the kinetics of NPs growth by the classical diffusion mechanism of Oswald ripening: the crystal size increases continuously with time. However, in the case of thoria, the kinetics of particle growth cannot be fitted by parabolic law. Moreover, the annealing time has no significant effect on particle size (Fig. 2). Recently, the so-called non-classical mechanism of oriented attachment was proposed to describe specific processes of nanocrystals growth. The growth of NPs by the mechanism of oriented attachment has been experimentally proven for many compounds, including $TiO_2$, $MnO_2$, ZnO, $Fe_2O_3$, $CeO_2$, etc. Using the HRTEM technique, Penn and Banfield [60] demonstrated the growth of $TiO_2$ particles by this mechanism under hydrothermal conditions. Ivanov et al.[61] suggested that the formation of larger $CeO_2$ crystallites as a result of high-temperature annealing also occurs according to the oriented attachment mechanism. The mechanism of oriented attachment could be successfully applied to describe thorium NPs growth during high-temperature annealing. The overall process could be described as follows: as a result of the synthesis and subsequent drying of the oxide samples, aggregated NPs are formed, which are randomly located in relation to one another, and thereby form a set of high- and low-angle boundaries. Each temperature increase promotes mutual orientation of crystallites, beginning from the particles with the least mutual disorientation; with a further increase in temperature, the more disoriented crystallites are forced to move. Upon self-organization of adjacent particles, so that they share a common crystallographic orientation, the joining of these particles at a planar interface occurs. Further growth of a new particle takes place through the surface diffusion process.

To confirm the stability of the local phase composition and morphology of the NPs during the annealing process, as well as to support XRD data, the sample obtained by annealing at 400 °C for 4 hours was studied by HRTEM and electron diffraction (Fig. 2 c,d). The electron diffraction pattern proves the stability of the cubic Fm3m $ThO_2$ phase is stable during annealing. The average size of $ThO_2$ nanocrystals annealed at 400 °C for 4 hours was established to be 4.8 ±0.9 nm.

## Size effect on $ThO_2$ crystal structure

The local structural parameters were determined by curve-fitting of the EXAFS χ-functions extracted from the raw experimental data. Fig. 3 shows the Fourier transform of the $k^3$-weighted EXAFS function of nanocrystalline $ThO_2$ samples with different particle sizes. As shown in Fig. 3, the radial distribution functions for all studied samples are almost identical. In the Fourier transforms, the first strong peaks at ~ 1.9 Å correspond to Th - O coordination, which in all the samples are quite similar in height and position. In contrast, the intensity of the Th-Th coordination peak at ~ 3.8 Å decreases strongly with decreasing particle diameter, and vanishes for the smallest sample. These data indicate a similar local atomic structure around the Th atom in all the samples, regardless of particle size and synthesis procedure.

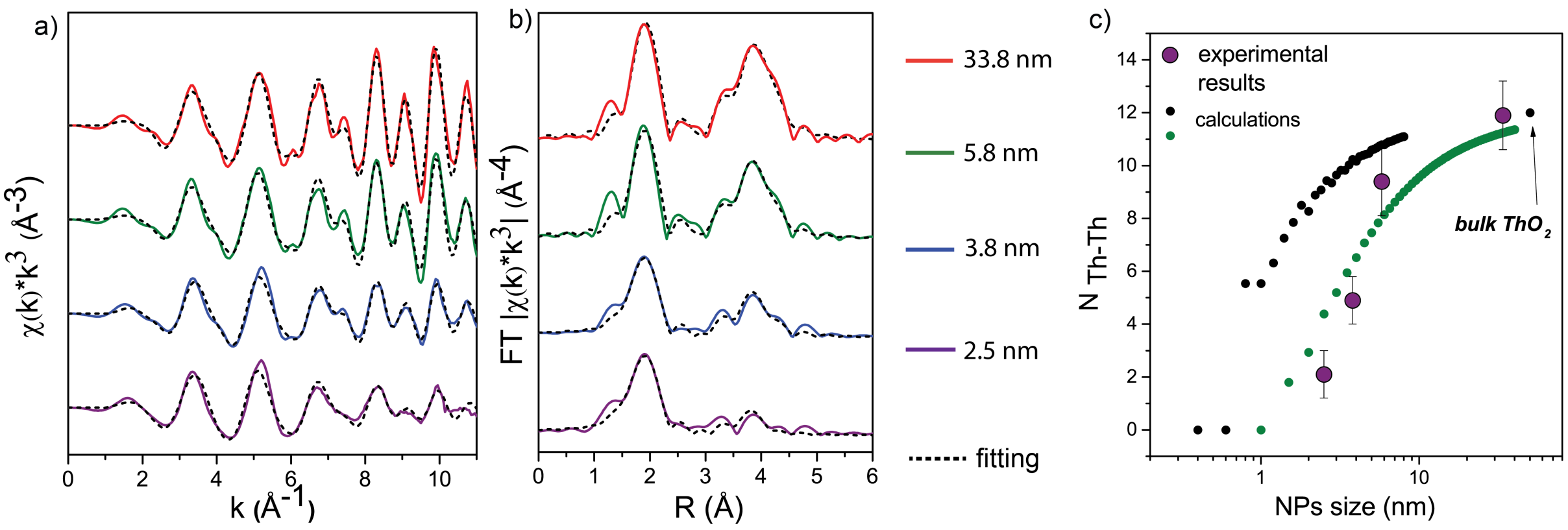

**Figure 3.** Th L3-EXAFS R-space fitting results for 2.5, 3.8, 5.8 and 33.8 nm $ThO_2$ samples: a) $k^3$-weighted $\chi(k)$ experimental functions (coloured solid lines) and corresponding fitting models (dashed lines); b) FT magnitude of EXAFS data (coloured solid lines), fit magnitude (dashed lines); c) Comparison of the size-dependent change of Th-Th coordination number in $ThO_2$ NPs, as determined by EXAFS vs. calculated values for $ThO_2$ particles of different size; black dots correspond to value of Th atoms in the second coordination shell in spherical NPs with a $ThO_2$ crystal structure; green dots correspond to calculated effective Th-Th coordination number according to equation (2).

Table 1 summarizes the EXAFS data. The first oxygen coordination shell in nanocrystalline oxides is typically similar to the bulk counterparts, because the under-coordinated metal atoms on the surface complete their coordination shell with hydroxyl groups or water molecules.[62] A more detailed analysis of the first coordination shell shows the increase of the Debay-Waller factor with particle size reduction due to some static dissorder on the surface of the particles.[63] The second and the further coordination shells in NPs are more structure-, size- and shape-sensitive. Because of the high surface to volume ratio, metal-metal coordination numbers and distances in the NPs are different from those of the bulk.[64] The same tendency is observed in $ThO_2$ samples: the mean Th-Th coordination numbers obtained from the EXAFS data fits are notably reduced in small NPs. Fig. 3c shows the experimentally determined Th-Th coordination numbers, together with the calculated values for spherical NPs, as a function of particle diameter. The values of coordination numbers have been calculated by counting Th atoms in the second coordination shell of Th in spherical NPs with a $ThO_2$ crystal structure. Interestingly, the coordination numbers determined by EXAFS for 2.5 nm and 3.8 nm $ThO_2$ samples are significantly lower, even taking into account measurement error. In turn, the estimated Th-Th coordination number for the larger particles is quite close to the theoretical calculation. The 33.8 nm $ThO_2$ sample had the same coordination number for the first Th-Th shell as the bulk material. To explain the difference between calculated coordination numbers and the results of EXAFS spectra fitting procedure, another structural model of NPs was proposed, namely "core-shell" $ThO_2$ NPs. Th atom was suggested to belong to the core of $ThO_2$ NP when it has 12 Th atoms in the second coordination shell; otherwise, Th atom was suggested to belong to the shell of $ThO_2$ NP. To calculated the effective Th-Th coordination number ($CN_{ef}$) we normalize coordination number of core Th atoms to the total amount according to the equation (2):

$$CN_{ef} = 12 \cdot \frac{N_{core}}{N_{total}} \qquad (2)$$

were $N_{core}$ – number of Th atoms in core and $N_{total}$ - total number of Th atoms in the NP. In this case, the calculated and EXAFS-derived coordination numbers match more accurately.

Extremely low values of metal-metal coordination numbers are typical for many oxide and chalcogenide NPs, including $TiO_2$,[65] CdS,[66] ZnS,[67] CdSe,[68] etc. In the smaller clusters, the vanishing metal-metal coordination shell could be explained by sufficiently large static bond distance fluctuations because of (i) the presence of non-equivalent clusters and (ii) the changing of the lattice parameter from bulk to surface in the structure of a single particle.

Table 1. Structural parameters obtained from the fitting of EXAFS spectra for 2.5, 3.8, 5.8 and 33.8 nm $ThO_2$ samples.

| Size of $ThO_2$ NPs, nm | Coordination number (CN) | Coordination shell | Neighbouring atomic distance (R), Å | Debye-Waller factor ($\sigma^2$), Å$^2$ |
|---|---|---|---|---|
| 2.5 | 8.8±1.6 | O1 | 2.42±0.02 | 0.011 |
| | 2.3±0.9 | Th | 3.95±0.02 | 0.005* |
| 3.8 | 7.9±1.3 | O1 | 2.41±0.02 | 0.009 |
| | 5.1±0.9 | Th | 3.96±0.02 | 0.005* |
| | 3.3±12.4 | O2 | 4.71±0.13 | 0.005 |
| 5.8 | 7.6±0.8 | O1 | 2.41±0.01 | 0.005 |
| | 9.1±1.3 | Th | 3.96±0.01 | 0.005* |
| | 10.3±11.8 | O2 | 4.65±0.10 | 0.005 |
| 33.8 | 8.0±0.5 | O1 | 2.41±0.01 | 0.005 |
| | 11.6±1.3 | Th | 3.97±0.01 | 0.005* |
| | 16.9±10.3 | O2 | 4.63±0.04 | 0.005 |

*- Fixed values

Fig. 4 shows the dependence of the $ThO_2$ unit cell parameter on particle size. The cell parameter values and particle sizes were calculated using TOPAS software, from diffraction patterns obtained using laboratory and synchrotron XRD (Fig. S3). The size of NPs was estimated from FWHM, using the Scherrer equation. A comparison of the particle sizes obtained by the Scherrer method and calculated using TOPAS software (taking account of micro-stresses in NPs' structure) is presented in Table S2.

According to the obtained dependence (Fig. 4), the unit cell value varies from 5.66 Å to 5.60 Å, (the latter value corresponds to the bulk material), when the particle size changes from 2.5 nm to 35 nm. The maximum change in the lattice parameter (1.1 % from the bulk material lattice parameter) was registered for 2.5 nm NPs. The obtained dependence can be adequately fitted using the power-law relation: $\Delta a = a - 5.59912 = 0.142 \cdot D^{-1.09}$, where $a$ is the unit cell parameter (nm) $\Delta a = a_{bulk} - a$.

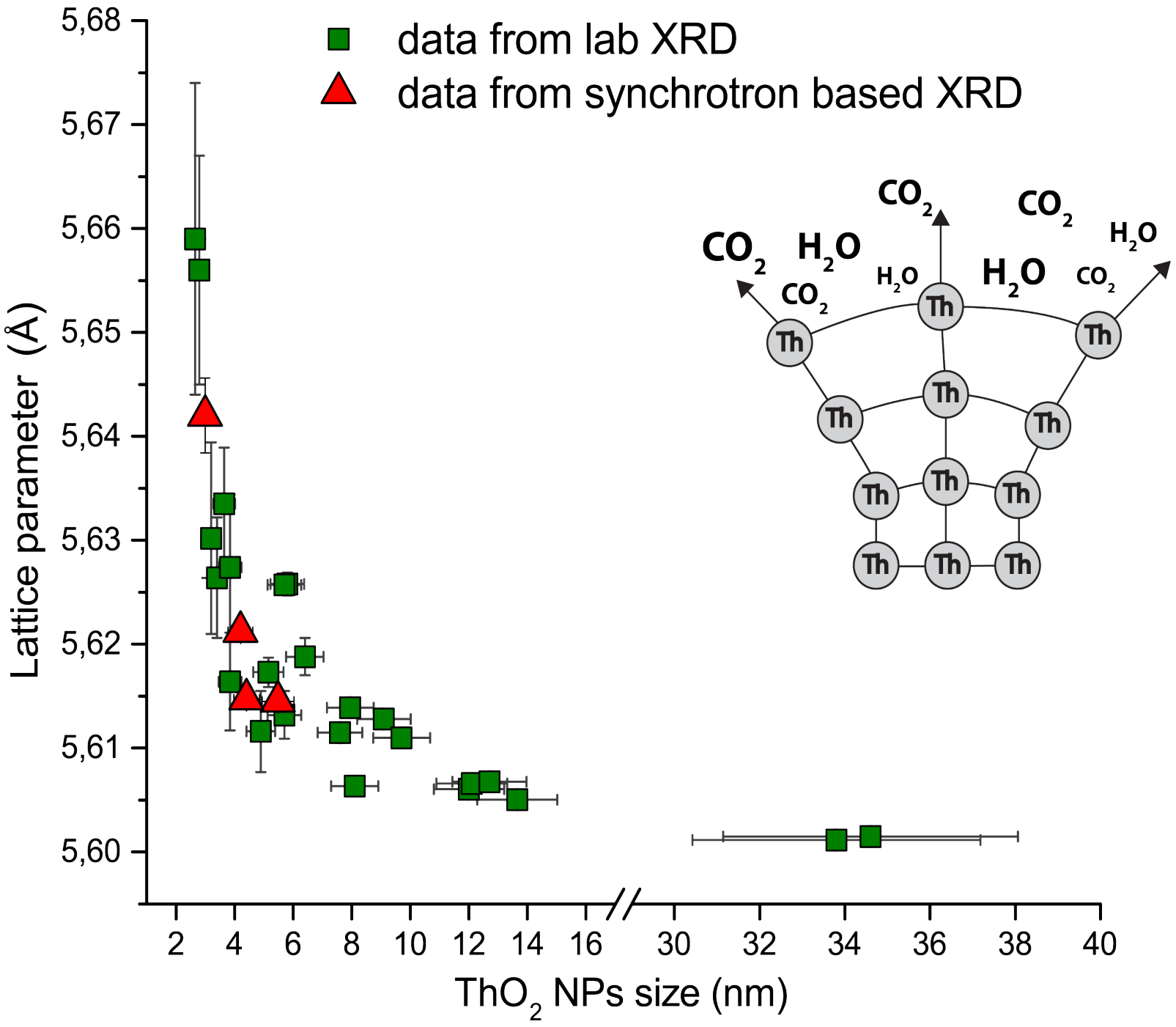

**Figure 4.** $ThO_2$ unit cell parameter dependence on particle size. Insert: schematic representation of the lattice expansion mechanism

The value of the unit cell parameter for NPs of various composition and size is an important fundamental property affecting their application in technological fields; it is of special importance for catalysts, and magnetic and optical materials. Several reports have been dedicated to the determination of unit cell parameter dependence on NPs' size for oxide NPs, including $CeO_2$, MgO, $Co_3O_4$, $Fe_3O_4$, $TiO_2$, etc. [69-74] In turn, the only available data concerning $ThO_2$ NPs' unit cell parameter values have been reported by Batuk et al. [28] on the basis of HRTEM analysis of a very limited set of NPs (2, 3 and 5 nm).

In earlier papers, the phenomenon of lattice expansion with NPs' decreasing size was considered to be an anomaly, as the first studies conducted on metals' NPs showed lattice compression instead of expansion. The most significant paper in the field, which included a large body of experimental data and theoretical calculations, has been published by Diehm et al. [75] They noticed that expansion of the crystal lattice with decreasing particle size is as common as its compression. According to Diem et al.,[75] there are three major theories for describing the phenomenon of lattice expansion: the "Madelung model",[76] the vacancy formation mechanism [77] and the theory of surface stress, the latter being the most common. According to ab initio calculations, the formation of various types of defects, including Frenkel pairs, oxygen or thorium vacancies, interstitial oxygen in the structure of thorium oxide is thermodynamically favoured [78-79]. Unfortunately, works deal with bulk thorium dioxide crystals. The formation of oxygen vacancies should lead to a change in the charge of the oxidation state of the neighbouring atom [77, 80]. However, in our recent paper [81], Th $L_3$ edge HERFD XAS spectra obtained for $ThO_2$ NPs indicated the presence of only thorium(IV) species (>95%).

All compounds (both metal and ionic materials) exhibit surface stress because of the different coordination of atoms on the surface and in the bulk. Compression or expansion of the crystal lattice with decreasing particle size depends on compound type: metal and halide NPs usually exhibit lattice contraction (positive surface stress), while oxides most often demonstrate lattice expansion (negative surface stress). The surface stress effect becomes more pronounced as the size of particles is reduced to nanometers, as the contribution of the surface atoms to the structural characteristics increases significantly.

The surface stress theory enables a qualitative description of the expansion of the crystal lattice in $ThO_2$ NPs. An oxide surface is highly charged because the ionic bonds break at the surface. Highly charged oxides readily adsorb $H_2O$ and $CO_2$ species under ambient conditions. Thus, the atoms on the surface become over-coordinated. According to fundamental principles, atom coordination number and bond length are related to each other: highly coordinated atoms tend to form longer bonds. According to the literature data,[82] when the coordination number of Th atoms changes from 6 to 12, its ionic radius increases from 0.94 to 1.21 Å. Therefore, the atoms on the $ThO_2$ surface exert a tensile influence on the bulk. Hydroxyl groups exist on the surface of thorium dioxide as a residue from the synthesis procedure or from the environment. In addition, the formation of carbonate surface species may occur.

To support this explanation for the effect of lattice expansion in $ThO_2$ NPs, and to prove experimentally the presence of $H_2O$ and $CO_2$ species on NPs' surface, TGA and MS measurements were performed for $ThO_2$ samples containing NPs of 2.5, 3.8 and 5.8 nm size. According to TGA analysis, the thermal decomposition process of the sample containing the smallest particles (2.5 nm) proceeds in three well-defined steps (Fig. 5, purple dashed curve). The first decomposition step (Step I) is observed at 25 – 160 °C, with a weight loss of 5 %. The second weight loss (Step II) occurs at 160 °C - 290 °C, and accounts for 2.8% of weight loss. The third decomposition step (Step III), with a weight loss of 4.9%, occurs at 290 – 600 °C. To obtain further insight into the thermal decomposition mechanism of $ThO_2$ NPs, TG-MS analysis was carried out. The mass spectra of $H_2O$ (MW 18) and $CO_2$ (MW 44) gases evolved during the thermal treatment of the samples are shown in Fig. 5 and Fig. S4, respectively. The body of data as a whole indicates that Step I corresponds to the release of physically adsorbed water only. The presence of water in the mass spectrum in the range of 160 – 290 °C indicates that the 2.5 nm sample was strongly hydrated, so its chemical composition should be presented as $ThO_2 \cdot xH_2O$, where x = 6. Step III corresponds to the removal of $CO_2$ from NPs' surface (Fig. S4a). TGA curves for 3.8 nm and 5.8 nm samples demonstrate much less total weight loss during the heating process: 6% and 2.7%, respectively. The mass spectra for 3.8 and 5.8 nm NPs also contain peaks corresponding to $H_2O$ and $CO_2$ evolution (Fig. S4 b, c). It should be noted that for 3.8 nm NPs the peak corresponding to water evolution is asymmetric and could be fitted by two peaks, corresponding to the water sorbed on the surface and water trapped in the bulk. Thus, the 3.8 nm $ThO_2$ sample could also be considered to have been hydrated. Unfortunately, since the TGA curve for this sample shows a continuous weight loss, which is also associated with $CO_2$ desorption, the precise determination of water content cannot be performed.

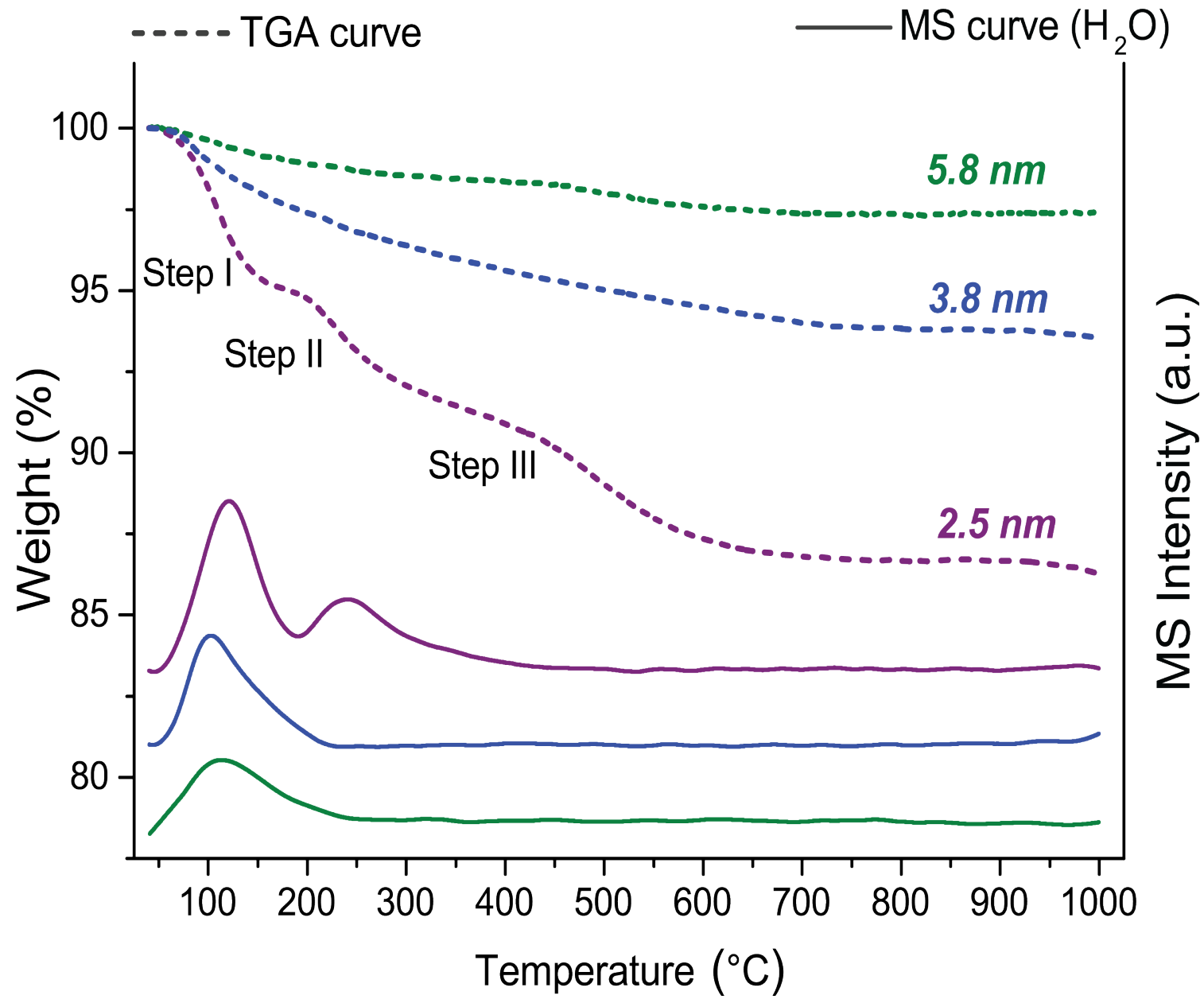

**Figure 5.** Thermal analysis data and the corresponding mass spectral data for $H_2O$ evolution from $ThO_2$ NPs of different sizes.

The influence of the local atomic environment on the lattice parameter values has been demonstrated previously for other stoichiometric oxides. According to Cimino et al.,[70] MgO particles that were kept in a vacuum showed a contraction of the lattice, while the particles exposed to the ambient atmosphere showed an expansion of the lattice. This effect was attributed to the formation of $Mg(OH)_2$ species on the surface of the particles caused by chemisorption of water. Diehm et al. computed surface stresses for oxide surfaces covered with several typical adsorbates, such as $H_2O$ and $CO_2$, to show the influence of environmental conditions on surface stress. For $TiO_2$, it was shown that the value of surface tension is sensitive to the presence of sorbents on the surface. According to TGA and MS data, $ThO_2$ samples synthesized in this work contained sorbed water and $CO_2$, and may also have contained chemically bound water molecules, which also contributes to lattice expansion.

## CONCLUSIONS

In this work, several soft chemical routes have been proposed for the production of thoria NPs in a wide range of average sizes, from 2.5 to 34 nm. By selecting the chemical precipitation conditions (thorium salt concentration, concentration and nature of the precipitants, etc.), the heat treatment method (mild drying, hydrothermal treatment, annealing) and the temperature of treatment, crystalline $ThO_2$ NPs could be produced in a strictly controlled manner. It has been shown that crystalline $ThO_2$ particles with a size of about 2 nm are formed during precipitation from thorium nitrate solution and subsequent gentle drying in air (T = 100 - 150 °C). Varying the concentration of thorium nitrate does not affect the NPs' size, while the use of sodium hydroxide as the precipitant leads to the formation of larger crystallites. The effect of NaOH as a mineralizer was also well pronounced during the hydrothermal treatment of NPs. Annealing of $ThO_2$ powders for 1–8 hours at 300–800 °C led to the growth of NPs from 3–4 to 30 nm, the size of the resulting NPs being dependent on temperature but not on annealing duration. For the first time, the dependence of the unit cell on $ThO_2$ particle size was obtained: the $ThO_2$ elementary unit showed a tendency to expand with decreasing particle size. By analogy with other metal oxides ($CeO_2$, $TiO_2$, MgO, etc.), the model of negative surface stress was used to describe the data obtained. The presence of surface hydroxyl and carbonate groups that have a tensile effect on the crystalline lattice has been proven experimentally by TG/MS measurements of selected samples. According to EXAFS analysis, $ThO_2$ particle size did not affect the coordination number in the first oxygen coordination sphere, while the metal–metal coordination number also depended significantly on the NPs' size.

## SUPPOROTING INFORMATION

XRD, ED, TGA/MS data and $ThO_2$ cell parameter values. This material is available free of charge via the Internet at http://pubs.acs.org/.

ACKNOWLEDGEMENTS. The study was supported by the Russian Foundation for Basic Research (project 16-33-60043mol_a_dk). K.O.K greatly acknowledge a support from the European Commission Council under ERC grant N759696. Experimental studies were partially performed on equipment acquired with funding from the Lomonosov Moscow State University Development Program. The authors are grateful to A.V. Garshev for his assistance in HRTEM measurements.